1

# The Carbon-Rich Gas in the β Pictoris Circumstellar Disk

Aki Roberge*, Paul D. Feldman†, Alycia J. Weinberger‡, Magali Deleuil,§ & Jean-Claude Bouret§

*Exoplanets & Stellar Astrophysics Laboratory, NASA Goddard Space Flight Center, Greenbelt, MD 20771, USA*

† *Department of Physics & Astronomy, Johns Hopkins University, Baltimore, MD 21218, USA*

‡ *Department of Terrestrial Magnetism, Carnegie Institution of Washington, Washington, DC 20015, USA*

§ *Laboratoire d'Astrophysique de Marseille, BP 8, F-13376 Marseille Cedex 12, France*

**The edge-on disk surrounding the nearby young star β Pictoris is the archetype of the "debris disks", which are composed of dust and gas produced by collisions and evaporation of planetesimals, analogues of Solar System comets and asteroids. These disks provide a window on the formation and early evolution of terrestrial planets. Previous observations of β Pic concluded that the disk gas has roughly solar abundances of elements[1], but this poses a problem because such gas should be rapidly blown away from the star, contrary to observations of a stable gas disk in Keplerian rotation[1,2]. Here we report the detection of singly and doubly ionized carbon (CII,**

**CIII) and neutral atomic oxygen (OI) gas in the β Pic disk; measurement of these abundant volatile species permits a much more complete gas inventory. Carbon is extremely overabundant relative to every other measured element. This appears to solve the problem of the stable gas disk, since the carbon overabundance should keep the gas disk in Keplerian rotation[3]. New questions arise, however, since the overabundance may indicate the gas is produced from material more carbon-rich than the expected Solar System analogues.**

β Pic is an A5 V star, indicating that it is approximately 1.8 times more massive than the Sun. It appears to have solar elemental abundances[4] and is 8 – 20 million years old[5]. Narrow atomic absorption lines were studied in the spectra of β Pic even before the star was known to have a dust disk[6]. The absorption features fall into one of two categories. Almost every line observed shows an unvarying narrow absorption at the radial velocity of the star. This component contains the bulk of the circumstellar gas and is called the stable component[1]. On the wings of most absorption lines are broad, variable absorption features, which are typically redshifted with respect to the star. These features arise from gas falling toward the star at high velocity, likely produced by vaporization of star-grazing planetesimals[7].

Various studies of the dynamics of the β Pic stable gas have shown that there is a serious gap in our understanding. The force of stellar radiation pressure on many of the observed atomic and ionic species should rapidly blow them away from the star. The very existence of gas at the velocity of the star (the bulk of the β Pic circumstellar gas) is puzzling. Workers were forced to postulate the existence of an undetected, dense hydrogen torus to brake some species through gas drag[1]. The problem got worse after the detection of resonantly scattered emission from gaseous species in the β Pic disk[8, 2]. These spatially resolved spectra show a gas disk in Keplerian rotation between 13 AU and a few hundred AU from the star. Analysis of the gas dynamics demonstrated that the some of the observed species should not be in Keplerian rotation, unless the total gas mass is much greater than the observed gas mass[2].

We used spectra from the *Far Ultraviolet Spectroscopic Explorer* (*FUSE*) to measure the volatile species CII, CIII, and most importantly, OI, in the circumstellar disk. These species are seen in absorption against broad stellar emission lines (Figure 1). Models of the absorption lines were compared to the data and the best absorption parameters (Table 1) determined through $\chi^2$ minimization (details of the observations, models, and analysis appear in the online Supplementary Information). Combined with a previous measurement of CI gas[9], the CII measurement completes the inventory of stable carbon gas in the disk. OI is the only ionization state of oxygen observed to date, and the only one likely to be abundant, given the high first ionization energy of oxygen. We also used archival *Hubble Space Telescope* Space Telescope Imaging Spectrograph (STIS) high-resolution far-UV spectra to search for absorption lines from a few species that have not yet been observed (CrI, NiI, PI, and PII), but none were detected.

Previous observations of the β Pic gas were primarily of metallic species[1]. The addition of our new measurements provides the most complete inventory of the gas in any debris disk. The β Pic elemental abundances, found by summing the abundances of the various ionization stages of each element, are shown in Table 2, along with the elemental abundances for the Sun[10], a carbonaceous chondrite meteorite[10], and comet Halley dust[11]. This table utilizes results from studies of the β Pic gas stretching over more than ten years[1, 2, 9, 12, 13, 14] (see the online Supplementary Information for details). A plot of the elemental abundances from Table 2 appears in Figure 2. It shows that the composition of the β Pic gas is dissimilar to the composition of all three comparison bodies. In particular, carbon is extremely overabundant relative to every other measured element. The β Pic C/O and C/Fe ratios are 18 and 16 times the solar value, respectively. The lithophile elements (e.g. Mg, Al) have roughly solar abundance relative to each other, as do the siderophile elements (e.g. Fe, Ni). However, the lithophile elements as a group seem slightly underabundant relative to the siderophile elements.

Neither oxygen nor carbon feels strong radiation pressure in the β Pic disk, since the star is relatively faint in the far-UV where the strong absorption lines of these species lie. By contrast, metals such as Na and Fe feel extremely strong radiation pressure and could be blown out of the system, producing apparent C and O overabundances relative to these elements. But carbon is overabundant relative to every other element, including ones that feel similarly weak radiation pressure, like oxygen. This leads us to suspect that the overabundance reflects the composition of the parent material, which would have to be more carbon-rich than the obvious Solar System analogue materials. The bulk composition of the Beta Pic planetesimals might be carbon-rich. There was a large increase in organics relative to water in the material excavated from comet Tempel 1 during its collision with Deep Impact[15]. Or the planetesimals might be selectively outgassing volatile carbonaceous compounds, trace amounts of which are still found in carbonaceous chondrite meteorites (J. Nuth, personal communication). There are probably many other possible explanations.

A new analysis of the gas dynamics in β Pic which took into account Coulomb forces between gaseous ions, neutral atoms, and charged dust showed that these interactions increase the effective collision cross-sections, dramatically reducing the need for additional unobserved braking gas[3]. But unless the dust grains are very highly charged, additional braking gas is still required. However, that analysis assumed solar abundances in the gas. If one considers our measured midplane abundances in conjunction with Coulomb forces, the problem may be solved. The ratio of the radiation and gravitational forces felt by an atom is the radiation pressure coefficient (β). The β-value for a particular species depends on the brightness of the central star at the wavelengths of the absorption lines of the species. Neutral elements that feel strong radiation pressure are quickly ionized before they can be blown away from the star; therefore, the species that must be braked are ionized. Coulomb interactions between the ions couple them together into a single fluid[3]. If the effective radiation pressure coefficient of the fluid ($\beta_{eff}$) is < 0.5, it will be bound to the star and the gas will be self-braking. Since carbon is abundant and moderately ionized in the β Pic disk, it is an important constituent of the ionic fluid. β Pic is faint in the far-UV, where the strong CI and CII absorption lines lie, so carbon feels negligible radiation pressure in the β Pic disk. Increasing the carbon abundance lowers $\beta_{eff}$. The ionic fluid will be self-braking ($\beta_{eff}$ < 0.5) if the carbon abundance is enhanced by a factor of 10 or more over the solar abundance[3]. Carbon in the β Pic stable gas is enhanced relative to oxygen by a factor of 18 over the solar C/O abundance, so additional unseen gas or highly charged grains are not needed to keep the gas disk in Keplerian rotation.

The reason why the gas composition is just what is required for Keplerian rotation is unknown. Answering this question will require similar observations and analyses to be done for many debris disks. However, *FUSE* spectra of another debris disk may provide a hint[16]. Circumstellar gas and dust have been detected around the B9 primary star of the σ Herculis binary system. In this case, the gas is moving away from the star. It has been suggested that the gas and dust are produced by collisions among planetesimals located about 20 AU from the star, where the orbits of small bodies become unstable in the binary system. A radiatively driven wind is generated as the high radiation pressure from this UV-bright star blows away the circumstellar gas. This might be the more common fate of debris gas around a high mass star, and perhaps would have occurred in the β Pic disk if not for the overabundance of carbon in its circumstellar gas. Of course, this does not explain how the carbon overabundance occurred in the first place. A great deal more observational and theoretical work is required to determine if the gas composition reflects the bulk composition of the parent bodies and then determine how such carbon-rich planetesimals could have formed in the β Pic protoplanetary disk.

**Supplementary Information** accompanies the paper on **www.nature.com/nature**.

We thank Don Lindler for recalibrating the archival STIS spectra used in this paper. The work at JHU was supported by NASA.

The authors declare no competing financial interests. Correspondence and requests for materials should be addressed to A.R. (e-mail: akir@milkyway.gsfc.nasa.gov).

**Table 1. Characteristics of the CII, OI, and CIII Absorption Lines**

| Species | Wavelength (Å) | $v_{FUSE}$ (km s$^{-1}$) | $v_{\text{heliocentric}}$ (km s$^{-1}$) | $b$ (km s$^{-1}$) | $N$ (atoms cm$^{-2}$) |
|---|---|---|---|---|---|
| C II | 1036.34 | $-5.0^{+5.8}_{-11.5}$ | 20 | $1.0^{+4.7}_{-1.0}$ | $1.0^{+1.7}_{-0.3} \times 10^{16}$ |
|  |  | $20.0^{+5.8}_{-9.3}$ | 45 | $61.0^{+11.8}_{-6.8}$ | $4.0^{+6.7}_{-1.0} \times 10^{14}$ |
| C II* | 1037.02 | $-5.0^{+5.8}_{-11.5}$ | 20 | $1.0^{+4.7}_{-1.0}$ | $1.0^{+1.2}_{-0.3} \times 10^{16}$ |
|  |  | $20.0^{+5.8}_{-9.3}$ | 45 | $61.0^{+11.8}_{-6.8}$ | $4.0^{+4.9}_{-1.2} \times 10^{14}$ |
| O I | 976.45 | $3.0^{+3.2}_{-2.8}$ | 20 | $14.0^{+3.6}_{-2.6}$ | $3.6^{+0.8}_{-0.6} \times 10^{15}$ |
| O I* | 977.96 | ⋯ | ⋯ | ⋯ | $\leq 2.6 \times 10^{15}$ |
| O I** | 978.62 | ⋯ | ⋯ | ⋯ | $\leq 8.7 \times 10^{14}$ |
| C III | 977.02 | $-7.0^{+2.1}_{-3.2}$ | 10 | $32.0^{+4.1}_{-3.6}$ | $4.5^{+0.2}_{-0.6} \times 10^{13}$ |

This table shows the best absorption parameters obtained through $\chi^2$ minimization; the models were compared to the 2002 *FUSE* spectra. Details of the models and fitting procedures appear in the online Supplementary Information. The errors given are the standard deviations of the parameters (±1σ) arising from statistical errors. In column 1, a * or ** indicates that the line arises from the first or second excited fine-structure energy level of the atom; otherwise, the line arises from the ground energy level. Column 2 shows the vacuum rest wavelengths of the absorption lines. In column 3, $v_{FUSE}$ is the velocity of the gas in the *FUSE* wavelength scale. *FUSE* does not carry an onboard calibration lamp, so the absolute wavelength scales of the spectra may have zero-point offsets. The accuracy of the relative wavelength calibration is about 8 km s$^{-1}$. Column 4 shows the heliocentric velocity of the gas, obtained by shifting the stable gas components to the stellar heliocentric velocity (20 km s$^{-1}$). Column 5 shows the best Doppler broadening parameters of the absorption components, which is a measure of the intrinsic width of the component. The last column shows the column densities in the various absorption components. The upper limits given for OI* and OI** were calculated assuming statistical population of the oxygen fine-structure levels (see Section 2.2 in the online Supplementary Information for an explanation). The smallest possible value for the total OI column density is the measured ground level OI column density (row 5). The largest possible value is the sum of the measured OI column density and the upper limits on the OI* and OI** column densities (rows 5, 6, and 7).

**Table 2. Elemental Abundances in the β Pic Stable Gas**

| Element | β Pic Column Density (atoms cm$^{-2}$) | Sun $N$(El) | CI Chondrite $N$(El) | Halley Dust $N$(El) |
|---|---|---|---|---|
| H  | $\lesssim$ few $\times 10^{19}$ | $2.88 \times 10^{10}$ | $5.50^{+0.67}_{-0.60} \times 10^{6}$ | $(1.09 \pm 0.21) \times 10^{7}$ |
| O  | $(3-8) \times 10^{15}$ | $1.41^{+0.17}_{-0.15} \times 10^{7}$ | $7.55^{+0.36}_{-0.34} \times 10^{6}$ | $(4.81 \pm 0.59) \times 10^{6}$ |
| C  | $5.0^{+2.3}_{-1.1} \times 10^{16}$ | $7.08^{+0.68}_{-0.62} \times 10^{6}$ | $7.72^{+1.14}_{-1.00} \times 10^{5}$ | $(4.40 \pm 0.89) \times 10^{6}$ |
| N  | $\cdots$ | $1.95^{+0.56}_{-0.44} \times 10^{6}$ | $5.54^{+0.97}_{-0.82} \times 10^{4}$ | $(2.27 \pm 0.76) \times 10^{5}$ |
| Mg | $\geq 2 \times 10^{13}$ | $1.00^{+0.05}_{-0.05} \times 10^{6}$ | $1.04^{+0.05}_{-0.05} \times 10^{6}$ | $5.41 \times 10^{5}$ |
| Si | $1 \times 10^{14}$ | $1.00^{+0.05}_{-0.05} \times 10^{6}$ | $1.00^{+0.05}_{-0.05} \times 10^{6}$ | $(1.00 \pm 0.10) \times 10^{6}$ |
| Fe | $(3.7 \pm 0.5) \times 10^{14}$ | $8.13^{+0.58}_{-0.54} \times 10^{5}$ | $8.63^{+0.62}_{-0.58} \times 10^{5}$ | $(2.81 \pm 0.49) \times 10^{5}$ |
| S  | $1 \times 10^{13}$ | $4.63^{+0.45}_{-0.41} \times 10^{5}$ | $4.45^{+0.43}_{-0.39} \times 10^{5}$ | $(3.89 \pm 1.24) \times 10^{5}$ |
| Al | $4.5 \times 10^{12}$ | $8.51^{+0.40}_{-0.38} \times 10^{4}$ | $8.31^{+0.39}_{-0.37} \times 10^{4}$ | $(3.68 \pm 0.92) \times 10^{4}$ |
| Ca | $2.6^{+5.3}_{-0.9} \times 10^{13}$ | $6.61^{+0.47}_{-0.44} \times 10^{4}$ | $6.00^{+0.43}_{-0.40} \times 10^{4}$ | $(3.41 \pm 1.03) \times 10^{4}$ |
| Na | $\geq 3 \times 10^{13}$ | $5.75^{+0.41}_{-0.38} \times 10^{4}$ | $5.75^{+0.41}_{-0.38} \times 10^{4}$ | $(5.41 \pm 3.24) \times 10^{4}$ |
| Ni | $1.5 \times 10^{13}$ | $4.78^{+0.34}_{-0.32} \times 10^{4}$ | $4.78^{+0.34}_{-0.32} \times 10^{4}$ | $(2.22 \pm 1.14) \times 10^{4}$ |
| Cr | $3.5 \times 10^{12}$ | $1.26^{+0.15}_{-0.14} \times 10^{4}$ | $1.31^{+0.16}_{-0.14} \times 10^{4}$ | $(4.86 \pm 1.08) \times 10^{3}$ |
| P  | $\leq 9.3 \times 10^{13}$ | $8.91^{+0.86}_{-0.78} \times 10^{3}$ | $7.83^{+0.76}_{-0.69} \times 10^{3}$ | $\cdots$ |
| Mn | $3 \times 10^{12}$ | $7.08^{+0.51}_{-0.47} \times 10^{3}$ | $9.17^{+0.66}_{-0.61} \times 10^{3}$ | $(2.70 \pm 1.08) \times 10^{3}$ |
| Zn | $(2-3) \times 10^{11}$ | $1.20^{+0.12}_{-0.11} \times 10^{3}$ | $1.25^{+0.12}_{-0.11} \times 10^{3}$ | $\cdots$ |

The total abundance for each element in the β Pic stable gas (column 2) was found by summing the column densities of the various ionization stages of the element, which appear in Table 2 of the online Supplementary Information. *N*(El) is the elemental abundance normalized to the number of Si atoms, where *N*(Si) = 10$^6$. The errors given are the standard deviations of the values (±1σ), where available.

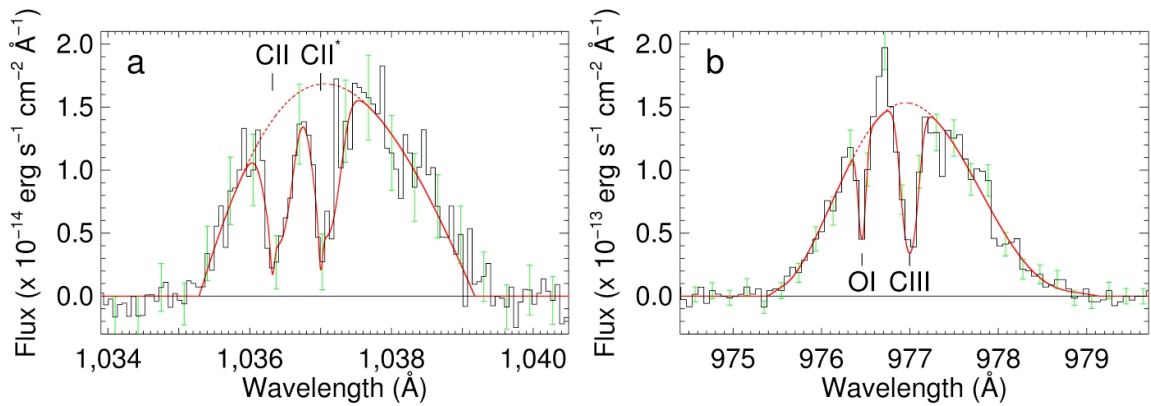

Figure 1. Circumstellar absorption lines in far-UV spectra of β Pic. β Pic was observed by *FUSE* once in 2002 and twice in 2001; the higher signal-to-noise spectra from 2002 are shown here. The error bars (shown in green) are the standard deviations of the flux values (±1σ). Details of the observations and analysis appear in the online Supplementary Information. Atomic absorption lines from CII, OI, and CIII are superimposed on broad emission lines arising in the chromosphere and transition zone of the star[17]. We model each emission feature with a polynomial (dashed red line). The total model (emission model times best absorption model from $\chi^2$ minimization) is overplotted with a solid red line. **(a)** The CII λλ1036.3,1037.0 absorption doublet superimposed on the OVI λ1038 emission line. There is also some CII emission blended with the blue wing of the OVI line[18]. The asymmetric shape of the CII absorption lines indicates there are two absorption components at different velocities. **(b)** The OI λ976.5 and CIII λ977.0 absorption lines superimposed on the CIII λ977 emission line. These lines are fitted with single absorption components. Underestimation of the OI column density due to saturation of the absorption line is unlikely, since the line is not strongly saturated. The interstellar contribution to the OI absorption line is negligible (see the Supplementary Information for details).

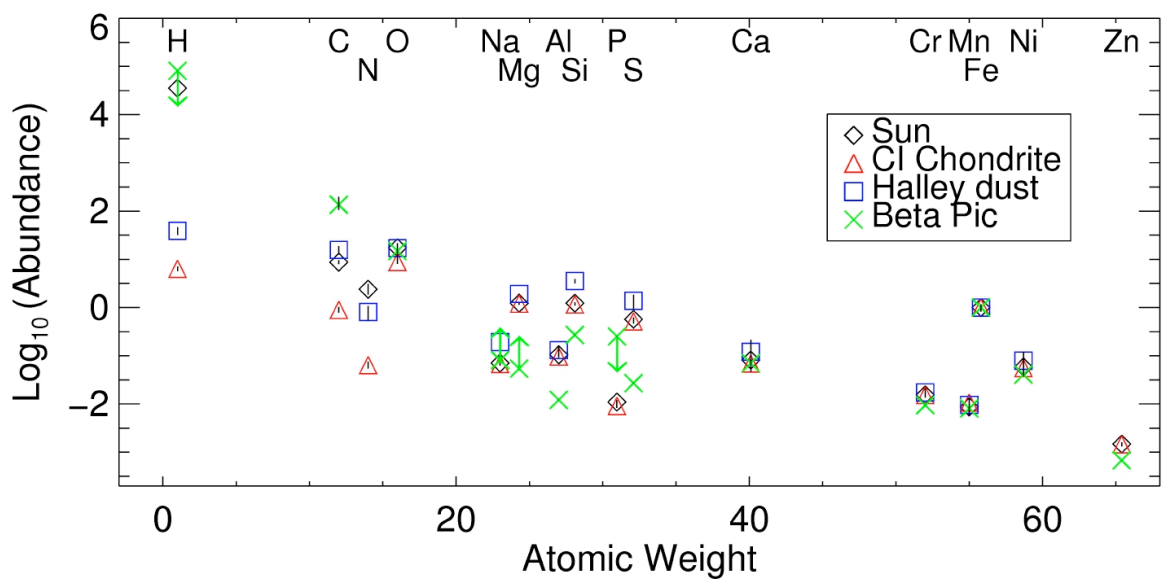

Figure 2. Bulk composition of β Pic stable circumstellar gas. The elemental abundances for the β Pic midplane gas are shown with green crosses, the Sun[10] with black diamonds, a carbonaceous chondrite meteorite[10] with red triangles, and comet Halley dust[11] with blue squares. The y-axis shows the logarithmic elemental abundance, normalized to Fe. The vertical lines on top of the data points show the standard deviations of the abundances (±1σ), where available. Upper and lower limits are indicated with arrows. Details of the measurements used in constructing this plot appear in the online Supplementary Information.